\documentclass[aps,prl,twocolumn,superscriptaddress,floatfix,amsmath,showpacs]{revtex4}

\usepackage{graphicx}

\begin{document}
\title{Formation of guided spin-wave bullets in ferrimagnetic film stripes}
\author{A.A.~Serga}
\affiliation{Fachbereich Physik and Forschungszentrum OPTIMAS, Technische Universit\"{a}t Kaiserslautern,
67663 Kaiserslautern, Germany}

\author{M.P.~Kostylev}
\altaffiliation{on leave from St.Petersburg Electrotechnical University, 197376, St.Petersburg, Russia}
\affiliation{School of Physics, The University of Western Australia,
Crawley WA 6009, Australia}

\author{B.~Hillebrands}
\affiliation{Fachbereich Physik and Forschungszentrum OPTIMAS, Technische Universit\"{a}t
Kaiserslautern, 67663 Kaiserslautern, Germany}


\begin{abstract}
The formation of quasi-2D nonlinear spin-wave eigenmodes in longitudinally magnetized stripes of a
ferrimagnetic film, so-called guided spin-wave bullets, was experimentally observed by using time- and
space-resolved Brillouin light scattering spectroscopy and confirmed by numerical simulation. They
represent stable spin-wave packets propagating along a waveguide structure, for which both transversal
instability and interaction with the side edges of the waveguide are important. The experiments and the
numerical simulation of the evolution of the spin-wave excitations show that the shape of the formed
packets and their behavior are strongly influenced by the confinement conditions. The discovery of these
modes demonstrates the existence of quasi-stable nonlinear solutions in the transition regime between
one-dimensional and two-dimensional wave packet propagation.
\end{abstract}


\maketitle

Stable two-dimensional localized nonlinear spin-wave excitations, so-called spin-wave bullets, have been
previously observed in thin ferrimagnetic films of yttrium-iron-garnet (YIG) magnetized along the
propagation direction
\cite{Linear_nonlinear_diffraction_dipolar_spin_waves_yttrium_iron_garnet_films_observed_space-_time-resolved_Brillouin_light_scattering,
Self-Generation_Two-Dimensional_Spin-Wave_Bullets,
Parametric_Generation_Forward_Phase-Conjugated_Spin-Wave_Bullets_Magnetic_Films}. These films were
practically unbounded in both in-plane directions compared to the transversal size of the spin-wave
packets and the wavelength of the carrier spin wave. In contrary, in a one-dimensional waveguide
structure, where the width is comparable to or smaller than the spin-wave wavelength, only quasi
one-dimensional nonlinear spin-wave objects were observed, which are spin-wave envelope solitons.
\cite{Collision_properties_quasi-one-dimensional_spin_wave_solitons_two-dimensional_spin_wave_bullets,
Backward-volume-wave_microwave-envelope_solitons_in_yttrium_iron_garnet_films}. Both for solitons and
bullets linear pulse spreading in the direction of propagation (so called longitudinal direction) due to
wave dispersion is compensated by longitudinal nonlinear compression. As for the transverse in-plane
direction, solitons are meant to have a stable transverse  distribution of their dynamic magnetization
coinciding with the profile of the lowest linear spin-wave eigenmode of the waveguide. On the contrary,
bullets show transverse nonlinear instability of attractive type which overcompensates transverse
diffraction broadening of the wave packet. The nonlinear compression would lead to a wave packet
collapse if the medium is lossless. Weak magnetic losses in a real magnetic film ensure a fine balance
of nonlinear narrowing of the packet and of its diffraction spreading for some distance of propagation
\cite{Linear_nonlinear_diffraction_dipolar_spin_waves_yttrium_iron_garnet_films_observed_space-_time-resolved_Brillouin_light_scattering}.
This results in a quasi-2D spatially localized bell-shaped waveform which is stable during the lifetime
of the bullet.

Here we report on the experimental observation of a stable spin-wave packet propagating along a
waveguide structure, for which both transversal instability and interaction with the side edges of the
film waveguide are crucial. The structure  can be considered as a transitional case between the 2D case
of a continuous film and the quasi-1D case of a narrow stripe. We show that in this case the nonlinear
wave dynamics is distinctively different from both the soliton and the bullet cases, and can be
considered as efficient coherent nonlinear mixing of spin-wave eigenmodes of the waveguide. Our theory shows that
this mixing is due to a specific interaction -- the pseudo-linear generation of higher order modes by
the fundamental one.

The experiment was carried out using a longitudinally magnetized long YIG film stripe of 2.5\,mm width and
7\,$\mu$m thickness. The magnetizing field was 1831\,Oe. The spin waves were excited by a rf magnetic field
created with a microstrip antenna of 25\,$\mu$m width placed across the stripe and driven by input microwave
current pulses 20\,ns in duration at a carrier frequency of 7.125\,GHz. The spatio-temporal behavior of the
traveling spin-wave packets was investigated by means of space- and time-resolved Brillouin light scattering
spectroscopy \cite{Physics_Reports}.

\begin{figure}[t]
\begin{center}
\scalebox{1}{\includegraphics[width=8.5 cm,clip]{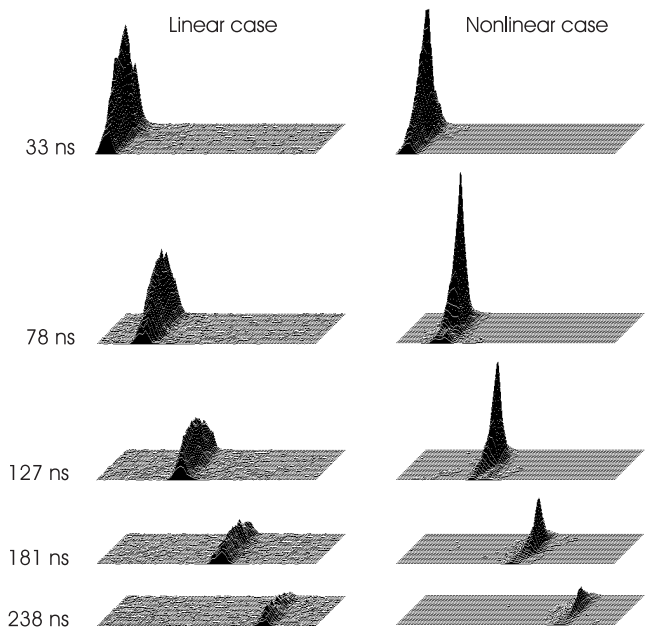}}
\end{center}
\vspace*{-0.4cm}\caption{Observation of linear propagation and of formation of the guided spin-wave
bullet in an YIG film stripe waveguide using time- and space-resolved Brillouin light scattering
spectroscopy. For parameters see main text.} \label{PulseProfiles}
\end{figure}

The results of our measurements are demonstrated in Fig.~\ref{PulseProfiles}. The panels show the
spatial distribution of the intensity of spin-wave packets at different times of their propagation from
the left to the right along the stripe waveguide. The left vertical set of diagrams corresponds to the
linear case. The input microwave power for this set is 20\,mW. The right set is for nonlinear
propagation. These data were collected applying an input driving power of 376\,mW.

Differences between these two cases are clearly seen. The linear spin-wave packet is characterized by a
transverse profile very similar to one half of the period of the sine function, while the cross section
of the nonlinear packet has a pronounced bell-like shape. Furthermore, the intensity of the linear
packet decays monotonically with time. That is obviously due to magnetic damping in the film. The
intensity of the nonlinear packet initially increases because of its strong transverse compression (see
the second diagram from the top in Fig.~\ref{PulseProfiles}), then gradually decays as the damping
overcomes the nonlinear compression for larger propagation times.

\begin{figure}[t]
\begin{center}
\scalebox{1}{\includegraphics[width=8.5 cm,clip]{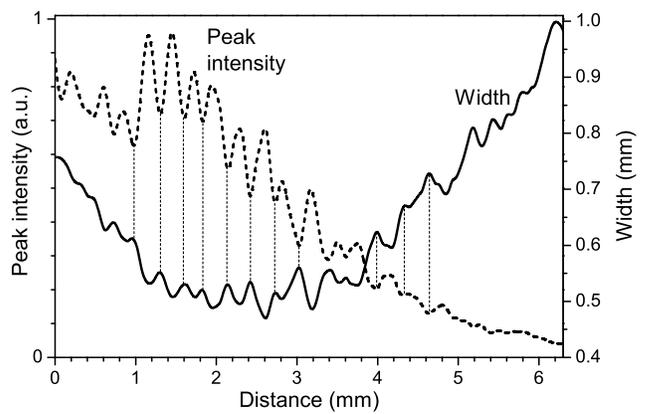}}
\end{center}
\vspace*{-0.4cm}\caption{Measured peak intensity and width for the nonlinear wave packet shown in the
right panels of Fig.~\ref{PulseProfiles} as a function of propagation path. The dashed vertical lines
show positions of local minima of intensity and of the local maxima of the packet width.} \label{Graph}
\end{figure}

Formation of a stable nonlinear spin-wave object is also evidenced in Fig.~\ref{Graph} which shows the
measured peak intensity of the nonlinear wave packet and its width (i.e. packet size in the transverse
direction) as a function of propagation path. At the beginning of the propagation path one clearly sees
an increase of the peak intensity on top of the exponential decay. The increase in peak intensity is due
to transverse compression of the packet, evidenced in the same figure by a decrease of the packet width.
Down the propagation path the width stabilizes, and in the range of distances $z$=1-4\,mm it does
practically not change, apart from small superimposed oscillations which will be discussed below.

Both figures provide a clear proof of development of transverse instability and bullet formation.
Interestingly, the 2D bell-like shape survives even at the end of the propagation path ($z>$4\,mm) when
the packet intensity has decreased more than ten times and the nonlinearity contribution to the
spin-wave dynamics has been considerably diminished.

Importantly, in Fig.~\ref{Graph} one observes superimposed oscillatory variations in the intensity and
in the packet width, both in the quasi-linear regime of the packet propagation ($z>4$\,mm) and in the
highly nonlinear regime $z<4$\,mm.  In the entire range the oscillations in intensity and in width are
in anti-phase to each other. This effect has not been observed in case of conventional bullet formation
\cite{Linear_nonlinear_diffraction_dipolar_spin_waves_yttrium_iron_garnet_films_observed_space-_time-resolved_Brillouin_light_scattering}.
On the contrary, a similar picture is usually observed for linear guided spin waves in narrow film
waveguides \cite{Snake,Demidov}, where it is explained as a beat of phase-correlated linear waveguide
width modes propagating at the same carrier frequency. This effect evidences the importance of the
influence of confinement on the nonlinear evolution of the packet transverse profile and suggests that
interaction of the linear eigenwaves of the waveguide -- the so-called width modes -- underlies
nonlinear wave dynamics. Therefore we term this nonlinear object a ``guided spin-wave bullet''.

The theoretical description of the observed phenomena is based on concepts developed in
Refs.~\cite{waveguide} and \cite{Demidov1}. The ingredients of the model are (i) effective dipole
pinning of magnetization, which results in a tri-linear interaction as the initial interaction for
development of the transverse instability of the wave packet profile, (ii) the width-mode group velocity
matching, and (iii) the nonlinear extension of spectrum of width modes. In Ref.~\cite{waveguide}, we
theoretically studied linear propagating eigenwaves of a magnetic stripe. The eigenwaves represent
guided modes with discrete transverse wavenumbers. For a stripe of rectangular cross-section the modes
are characterized by a standing-wave type dynamic magnetization distribution across the stripe
cross-section and by a monochromatic propagating wave with the longitudinal wavenumber $k_z$ along the
stripe. Importantly, for stripes with a large ratio $p$ of width to thickness the thickness distribution
of the dynamic magnetization is practically homogeneous, whereas in the direction of the stripe width
the standing spin waves possess considerably decreased amplitudes at the edges due to dynamic
demagnetization effects \cite{Guslienko}. Previous calculations (see Fig.~3 in \cite{waveguide}) show
that the assumption of a totally pinned magnetization at the stripe edges results in good approximation
for the transverse profiles of the propagating eigenmodes. Adopting the assumption of totally pinned
edge spins, the transverse profile of the dynamic magnetization is described by an integer number of
half-periods $n$ of the sine function with the lowest transverse wavenumber $k_y \equiv k_n$ being
$k_{n=1}=1 \cdot \pi/w$, where $w$ is the stripe width. For the aspect ratio of our waveguide, $p=357$,
this works with good accuracy.

As one sees from Fig.~\ref{PulseProfiles}, this conclusion is in a good agreement with the packet
profiles measured in the linear regime. The nonlinear dynamics is now described by a theory which is
analogous to the one developed in Ref.~\cite{Demidov1}. We assume that the dynamic pinning of
magnetization is conserved in the weakly nonlinear regime. Then the transverse evolution of the
nonlinear packet can be considered as interaction of linear eigenmodes which are pinned at the stripe
edges. Indeed, in Fig.~\ref{PulseProfiles} one clearly sees that in the nonlinear regime the dynamic
magnetization at the stripe edges practically vanishes. Then the description for the propagating modes
results in an evolutional equation for the spin-wave precession angle $\phi$ which reads:
\begin{eqnarray}
i\partial/\partial t F_{n,k} [\phi(y,z)] +(\omega_{n,k}+i\eta -\omega)F_{n,k}[\phi(y,z)] \\ \nonumber +T F_{n,k}
[\mid \phi(y,z) \mid^2  \phi(y,z)]=f_{n,k,t}.
\end{eqnarray}

In this expression $k\equiv k_z$, $\omega_{n,k}$ is the eigenfrequency of the $n$-th guided mode for the
longitudinal wavenumber $k$, $\eta$ is the relaxation frequency for the film, and $\omega$ is the
carrier frequency of the microwave signal $f(y,z,t)\exp(i \omega t)$ applied at the vicinity of $z$=0
which excites the input spin-wave packet, and $T$ is the nonlinear coefficient analogous to the
nonlinear coefficient of the spin-wave version of the Nonlinear Schr\"{o}dinger Equation
\cite{ZvezdinPopkov}. The operation $F_{n,k}$ denotes a 2D Fourier transform. It is the discrete sine
transform with the basis functions $\sin(k_n y)$ in the direction of stripe width $y$, and is the
Fourier integral over continuous wavenumbers $k$ with the basis functions $\exp(ikz)$ along the stripe.

The expression Eq.~(1) can be easily transformed into a system of dynamic equations for amplitudes of
guided modes $\phi_{n}(z,t)=F_{n}[\phi(y,z,t)]$ coupled by the four-wave nonlinear interaction. The analysis of
the system shows that the formation of the two-dimensional waveform can be considered as an extension of
the spectrum of the width modes. The partial waveforms carried by the individual width modes have the
same carrier frequencies equal to that of the external excitation signal $\omega$ and the carrier wave
numbers which satisfy the dispersion relations for the modes $\omega_{n,k}=\omega$. In the linear regime
all the modes are independent. In the nonlinear (high amplitude) regime the width modes become mutually
coupled which ensures intermodal coherent energy transfer.

The efficiency of mode coupling in the pulse regime depends on two major factors: the mode group
velocity matching and the type of nonlinear interaction. The geometry of a relatively wide stripe is
very favorable for having maximum contributions from both.

Let us first discuss the type of nonlinear mode interaction. The spin-wave packet immediately after
having been launched into the stripe is carried by the lowest (fundamental) width mode ($n=1$)
$\phi_{1}(z,t)$. Therefore it is necessary to consider the nonlinear interaction of higher-order width
modes with this particular mode.

The interactions of the fundamental width mode with all even modes is not important for symmetry
reasons. The nonlinear interaction of modes of the same type of symmetry is described by the parametric
term as well as by an additional pseudo-linear (tri-linear) term. The parametric interaction is of
convective instability type and is the same parametric instability which triggers formation of
conventional bullets in continuous films. This conventional process is described by a pair of
complex-conjugated equations with a parametric term proportional to the square of amplitude of the
pumping wave (see e.g. Eqs. (4) and (5) in \cite{induced modulational instability}). In this interaction
the packet carried by the fundamental mode plays the role of the pumping wave. Its energy is transferred
to the partial waveforms carried by the higher-order width modes. There is a threshold associated with
this parametric process due to natural damping in the medium. In the waveguide structure the threshold
is of the same order of magnitude as the modulation instability threshold in continuous films (see e.g.
Eq.~(10) in
Ref.~\cite{Linear_nonlinear_diffraction_dipolar_spin_waves_yttrium_iron_garnet_films_observed_space-_time-resolved_Brillouin_light_scattering}).
Furthermore, an initial perturbation in the form of non-vanishing amplitude of a higher-order mode is
needed to start the parametric amplification. This perturbation usually is provided by thermal
excitation. Therefore an amplified higher-order mode, which is group velocity matched with the pumping
wave, needs a large distance of propagation in order to reach a noticeable level. The energy of the
pumping wave decreases down the propagation path due to losses in the medium. If the parametric
amplification gain is small because of small supercriticality the higher-order mode cannot reach an
amplitude comparable with that of the fundamental mode before the amplitude of the pumping wave packet
carried by the fundamental mode falls below the threshold of parametric instability and the gain ceases.

What distinguishes the confined waveguide geometry is that the nonlinear mixing starts as a
pseudo-linear (tri-linear) interaction of the fundamental with the next-order symmetric mode which is
the third mode $\phi_{3}(z,t)$:  $\partial\phi_{3}/\partial t + v_{3} \partial\phi_{3}/\partial z
+i\omega^{\textrm{nl}}_{3}\phi_{3}=S_{13}(z,t)$. Here $v_{3}$ is the group velocity of the third mode,
$\omega^{\textrm{nl}}_{3}$ is its nonlinear frequency shift, and $S_{13}$ is the tri-linear
inhomogeneous term. This term has the form of a linear source of excitation with amplitude proportional
to $\left|\phi_{1}(z,t)\right|^3$ moving with the group velocity of the fundamental mode. The presence
of this pseudo-linear interaction at the early stage of the bullet formation is entirely due to the
effective dipolar pinning of the magnetization at the stripe edges. If the edge spins were unpinned, the
interaction of all the width modes would be purely parametric. The pseudo-linear excitation introduced
by this term into the dynamic equations is a threshold-free process, so it works even below the
threshold of parametric instability. Furthermore, in contrast to the parametric process this process
does not require non-zero initial amplitude of the amplified waveform to start the amplitude growth.
This mechanism ensures rapid growth of the symmetric $n = 3$ mode driven by the $n = 1$ mode  up to the
level where the parametric mechanism starts to work efficiently. After that the fundamental mode jointly
with the $n = 3$ mode is capable to rapidly generate a large set of modes with yet higher odd numbers
$n$ through both pseudo-linear and parametric mechanisms.

\begin{figure}[t]
\begin{center}
\scalebox{1}{\includegraphics[width=8.5 cm,clip]{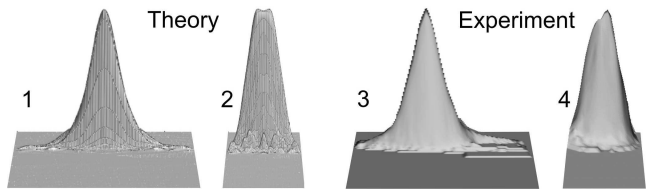}}
\end{center}
\vspace*{-0.4cm}\caption{Lateral shapes of the nonlinear SW packets. 1 and 2 -- theoretical results calculated
for the ferrite stripes of width of 2.5\,mm and 1\,mm, respectively. 3 and 4 -- experimental profiles observed in
YIG waveguides of width of 2.5\,mm and 1\,mm, respectively. 1 and 3: bullets. 2 and 4: solitons.}
\label{PulseProfiles1}
\end{figure}

Let us now discuss the importance of the mode group velocity matching. Our theory shows that the
efficiency of both nonlinear interaction mechanisms (parametric and tri-linear) strongly depends on the
group velocity difference of interacting modes and the initial length of the nonlinear packet. In wider
stripes the group velocities of the width modes are closer to each other. As the nonlinearity is of
attractive type, nonlinear corrections to the group velocities partially compensate for the group
velocity mismatch. As a result the nonlinearly generated higher-order partial waveforms remain for some
time within the pump packet. Again, due to the attractive character of the nonlinearity the modes have
initial phases such as their transverse profiles are summed up constructively in the middle of the
stripe width and a bullet-like total wave packet is formed as confirmed by direct numerical solution of
Eq.(1) shown in Fig.~\ref{PulseProfiles1}.

In the narrower stripes (1\,mm in width as in Fig.~\ref{PulseProfiles1}, panel~2) the mode group
velocity difference is larger and cannot be compensated by the nonlinear corrections to the group
velocities for the same initial intensity of the wave packet. Thus the group velocity matching is not
ensured. As a result the nonlinearly generated higher-order modes leave the area of the interaction
before they reach significant amplitudes. For the same length of the initial packet $\phi_{1}(t=0,z)$
the extension of the spectrum of width modes does not occur. The waveform $\phi_{1}(t,z)$ remains
unaffected by the interactions with the higher-order modes. It undergoes only a nonlinear longitudinal
compression and forms a quasi-1D wave packet -- the spin-wave envelope soliton -- which has a stable
sine-like profile in the $y$-direction.

The excellent agreement of the simulation results with the experimental data shown in
Fig.~\ref{PulseProfiles1} provides evidence for the validity of the developed theory.

In conclusion, the formation of quasi-2D nonlinear localized wave packets -- guided spin-wave bullets --
was studied in spin-wave waveguides. Our experimental and theoretical investigations show that formation
of these stable nonlinear objects is strongly affected by the transverse confinement of the medium. A
specific magnetostatic effect -- the effective dipole pinning of the magnetization at the edges of the
stripe, the width-mode group velocity matching of different discrete waveguide modes, and the extension
of the width mode spectrum due to nonlinear mode-mode energy transfer are essential for the nonlinear
evolution of the initial spin-wave excitation. Both the experimentally detected properties of the
evolution and the theoretically revealed mechanism of formation show that the observed nonlinear wave
packets can be treated as ``guided spin-wave bullets'' which are specific for laterally confined
magnetic films.

Support by the DFG within the SFB/TRR~49, the Australian Research Council, the University of Western Australia,
and the Russian Foundation for Basic Research is gratefully acknowledged.

\end{document}